\documentclass[aps,pra,superscriptaddress,12pt,tightenlines,nofootinbib]{revtex4}
\usepackage{amsmath,amsthm,graphicx,amssymb,verbatim,listings}
\usepackage[T1]{fontenc}     
\usepackage{lmodern}         
\usepackage[ pdftex, plainpages = false, pdfpagelabels,
                 pdfpagelayout = useoutlines,
                 bookmarks,
                 bookmarksopen = true,
                 bookmarksnumbered = true,
                 breaklinks = true,
                 linktocpage=all,
                 pagebackref=false,
                 colorlinks = true,
                 linkcolor = BrickRed,
                 urlcolor  = blue,
                 citecolor = BrickRed,
                 anchorcolor = green,
                 hyperindex = true,
                 hyperfigures
                 ]{hyperref}
\usepackage[usenames, dvipsnames]{xcolor}
\usepackage{xifthen} 

\newcommand{\arxiv}[2][]{\ifthenelse{\isempty{#1}}{\href{http://arxiv.org/abs/#2}{{\tt arXiv:\allowbreak{}#2}}} {\href{http://arxiv.org/abs/#2}{{\tt arXiv:\allowbreak{}#2 [#1]}}}}

\newcommand{\booktitle}{\textsl}
\newcommand{\hrefdoi}[2]{\href{https://dx.doi.org/#1}{#2}}

\begin{document}

\title{The Stock Market Has Grown Unstable Since February 2018}

\author{Blake C.\ Stacey}
\affiliation{\href{http://www.physics.umb.edu/Research/QBism/}{QBism Group}, Physics Department,\\ University of Massachusetts Boston}
\author{Yaneer Bar-\!Yam}
\affiliation{\href{http://necsi.edu}{New England Complex Systems Institute}}

\date{1 June 2018}

\maketitle

On the fifth of February, 2018, the Dow Jones Industrial Average dropped 1,175.21 points, the largest single-day fall in history in raw point terms.  This followed a 666-point loss on the second, and another drop of over a thousand points occurred three days later.  It is natural to ask whether these events indicate a transition to a new regime of market behavior, particularly given the dramatic fluctuations --- both gains and losses --- in the weeks since.  To illuminate this matter, we can apply a model grounded in the science of complex systems~\cite{deAguiar:2011, Stacey:2015, Allen:2017}, a model that demonstrated considerable success at unraveling the stock-market dynamics from the 1980s through the 2000s~\cite{Harmon:2010}.  By using large-scale comovement of stock prices as an early indicator of unhealthy market dynamics, this work found that abrupt drops in a certain parameter $U$ provide an early warning of single-day panics and economic crises~\cite{Harmon:2010}.  Decreases in $U$ indicate regimes of ``high co-movement'', a market behavior that is not the same as volatility, though market volatility can be a component of co-movement.  Applying the same analysis to stock-price data from the beginning of 2016 until now, we find that the $U$ value for the period since 5 February is significantly lower than for the period before.  This decrease entered the ``danger zone'' in the last week of May, 2018.

\begin{figure}[h]
  \includegraphics[width=14cm]{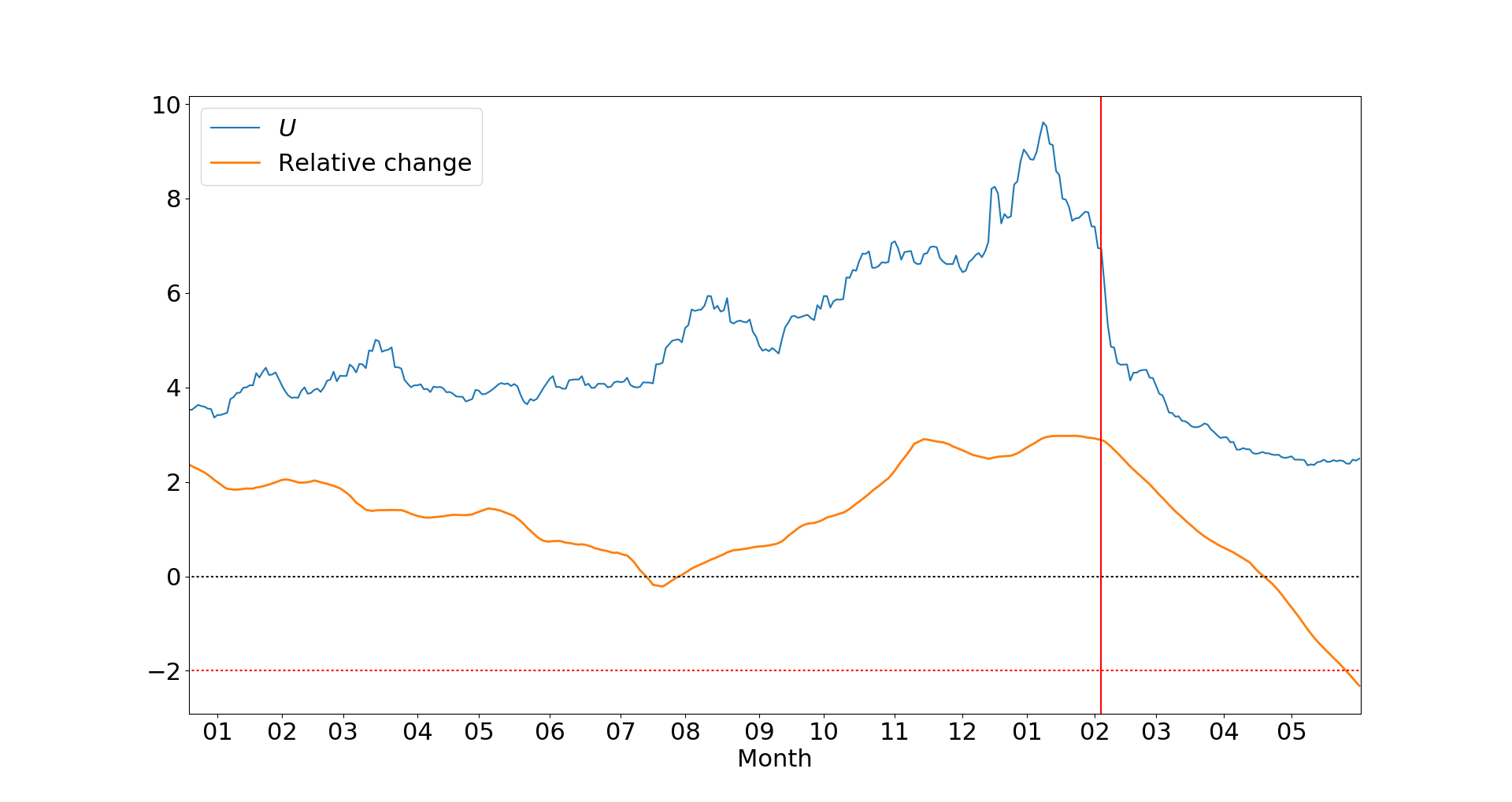}
  \caption{The market-dynamics parameter $U$ as a function of time, and its relative change over time, computed following the methods of prior work~\cite{Harmon:2010}, with a window of 81 trading days (the timespan from 5 February through 31 May) instead of 252 (one year). Higher values of $U$ indicate a market that is less prone to panic behavior. The vertical line indicates 5 February 2018. The relative change line crossing $-2$ is the indicator of unhealthy market mimicry found by~\cite{Harmon:2010}. Stock prices for the same set of tickers used in~\cite{Harmon:2010} were obtained from the EOD dataset released by Quandl~\cite{Quandl}.}
\end{figure}

Individual changes in market indicators like the Dow Jones average are typically attributed to dramatic news events.  For example, the loss of 572 points on 6 April 2018 was attributed to fears of a trade war between the United States and China~\cite{Meyersohn:2018}, and a drop of 391 points on 29 May was blamed on political turmoil in Italy~\cite{Liu:2018}.  However, the severity of the change provoked by any given external event depends upon the health of the market and its ability to react to stress.  The high levels of market mimicry observed in recent weeks indicate a complex system in poor health, which makes recent moves to loosen financial regulations~\cite{Rappeport:2018, Pyke:2018} particularly worrisome.  Regulations can have different effects, but they generally tend to reduce uncertainty in the market, thereby reducing volatility and risk. Moreover, some regulations have been instituted with the express intent of limiting instability. The stability of the stock market may already be lower compared to historical conditions, due to the weakening of the uptick rule to one based upon circuit breakers~\cite{uptick1, uptick2}.

Closely following the lead of prior work~\cite{Harmon:2010}, one would conclude that the market is now primed for a crisis to occur within the next year, a crisis that would manifest as a large single-day \emph{percentage} drop of the Dow Jones average. The market drop on 5 February that prompted this analysis does not itself qualify as a crash, since while it was the record-setter in raw point terms it only represented 4.6\% of the overall index value. Crashes correspond to drops in the 5\%--8\% range. This analysis raises the possibility of a loss of this magnitude, which could manifest as a single-day drop of 1,200 to 1,900 points.

We caution against taking this prediction as definitive, as the time window used for the analysis here is two months shorter than the shortest considered in~\cite{Harmon:2010}, and this may impair the model's predictive capability. The more robust conclusion to be drawn is that \emph{if} a serious event occurs, the market is in a poor state to deal with it.


\begin{thebibliography}{9}
\bibitem{deAguiar:2011} M.\ A.\ M.\ de Aguiar and Y.\ Bar-\!Yam, ``\hrefdoi{10.1103/PhysRevE.84.031901}{Moran model as a dynamical process on networks and its implications for neutral speciation},'' \booktitle{Physical Review E} \textbf{84}, 3 (2011), 031901, \arxiv{1012.3913}.
  
\bibitem{Stacey:2015} B.\ C.\ Stacey, \booktitle{Multiscale Structure in Eco-Evolutionary Dynamics.} PhD thesis, Brandeis University, 2015. \arxiv{1509.02958}.

\bibitem{Allen:2017} B.\ Allen, B.\ C.\ Stacey and Y.\ Bar-\!Yam, ``\hrefdoi{10.3390/e19060273}{Multiscale information theory and the marginal utility of information},'' \booktitle{Entropy} \textbf{19}, 6 (2017), 273.
  
\bibitem{Harmon:2010} D.\ Harmon, M.\ Lagi, M.\ A.\ M.\ de Aguiar, D.\ D.\ Chinellato, D.\ Braha, I.\ R.\ Epstein and Y.\ Bar-\!Yam, ``\hrefdoi{10.1371/journal.pone.0131871}{Anticipating economic market crises using measures of collective panic},'' \booktitle{PLOS One} \textbf{10}, 7 (2015), e0131871.

\bibitem{Quandl} ``End of Day US Stock Prices.'' Quandl (subscription required). \url{https://www.quandl.com/data/EOD-End-of-Day-US-Stock-Prices}.

\bibitem{Meyersohn:2018} N.\ Meyersohn, ``Dow tumbles 572 points as trade war fears pummel stocks,'' \booktitle{CNN Money} (6 April 2018). Downloaded from \url{http://money.cnn.com/2018/04/06/investing/stock-market-dow-jones-trade-war-china/index.html}.
  
\bibitem{Liu:2018} E.\ Liu, ``Markets Now: Dow Drops 391 Points as Italy Punches Above Its Weight,'' \booktitle{Barron's} (29 May 2018). Downloaded from \url{https://www.barrons.com/articles/markets-now-dow-drops-197-points-as-euro-crisis-returns-with-a-vengeance-1527593179}.

\bibitem{Rappeport:2018} A.\ Rappeport and E.\ Flitter, ``Congress approves first big Dodd--Frank rollback,''\\ \booktitle{The New York Times} (22 May 2018). \url{https://www.nytimes.com/2018/05/22/business/congress-passes-dodd-frank-rollback-for-smaller-banks.html}.

\bibitem{Pyke:2018} A.\ Pyke, ``Federal Reserve grants Wall Street's biggest wish in new push to dismantle Dodd--Frank centerpiece,'' \booktitle{ThinkProgress} (30 May 2018).\\ \url{https://thinkprogress.org/volcker-rule-dismantle-fed-a989d1388b66/}

\bibitem{uptick1} ``SEC Approves Short Selling Restrictions.''
  \booktitle{Securities and Exchange Commission} (24 February
  2010). \url{https://www.sec.gov/news/press/2010/2010-26.htm}.

  \bibitem{uptick2} Y.\ Bar-\!Yam, D.\ Harmon, V.\ Misra and
    J.\ Ornstein,
    ``\href{http://www.necsi.edu/research/economics/uptickmarket.html}{Regulation
      of Short Selling: The Uptick Rule and Market Stability}.''
    Report presented at the SEC (22 February 2010).
  
\end{thebibliography}
\end{document}